\documentclass[preprint,5p,twocolumn]{elsarticle}
\usepackage{graphicx}
\usepackage{comment}
\usepackage{siunitx}
\usepackage{gensymb}
\usepackage{booktabs}
\usepackage{color}
\usepackage{amsmath,amsfonts,amssymb,amsbsy}
\usepackage[toc,page]{appendix}
\biboptions{sort&compress}
\usepackage{cprotect}
\newcommand\blfootnote[1]{%
  \begingroup
  \renewcommand\thefootnote{}\footnote{#1}%
  \addtocounter{footnote}{-1}%
  \endgroup
}

\begin{document}
\begin{frontmatter}

\title{The effect of pressure pulsing on the mechanical dewatering of nanofiber suspensions}
%Inter-particle friction in acceleration of submerged granular flows\\
%Increasing friction increases acceleration of submerged granular flows}

\author[1]{Marko Korhonen\corref{cor1}}\ead{marko.korhonen@aalto.fi}
\author[1]{Antti Puisto}\ead{antti.puisto@aalto.fi}
\author[1]{Mikko Alava}\ead{mikko.alava@aalto.fi}
\author[2]{Thaddeus Maloney}\ead{thaddeus.maloney@aalto.fi}
\cortext[cor1]{Corresponding author}
\address[1]{Department of Applied Physics, Aalto University,  P.O. Box 11100 FI-00076 AALTO, Finland}
\address[2]{Department of Bioproducts and Biosystems, Aalto University, P.O. Box 16300 FI-00076 AALTO, Finland}
%\author{Antti Puisto}
%\author{Mikko Alava}
%\affiliation{Department of Applied Physics, Aalto University,  P.O. Box 11100 FI-00076 AALTO, Finland}
%\affiliation{Department of Bioproducts and Biosystems, Aalto University, PO Box 16300
%FI-00076 AALTO Finland}

%\date{\today}
\begin{abstract}
Dewatering processes are invariably encountered in the chemical manufacturing and processing of various bioproducts. In this study, Computational Fluid Mechanics (CFD) simulations and theory are utilized to model and optimize the dewatering of commercial nanofiber suspensions. The CFD simulations are based on the volume-averaged Navier-Stokes equations, while the analytical model is deduced from the empirical Darcy's law for dewatering flows. The results are successfully compared to experimental data on commercial cellulose suspensions obtained with a Dynamic Drainage Analyzer (DDA).
Both the CFD simulations and the analytical model capture the dewatering flow profiles of the commercial suspensions in an experiment using a constant pressure profile. However, a temporally varying pressure profile offers a superior dewatering performance, as indicated by both the simulations and the analytical model. Finally, the analytical model also predicts an optimized number of pressure pulses, minimizing the time required to completely dewater the suspension.
\end{abstract}


%Dewatering processes associated with fibrous suspensions are encountered in many industrial %settings, including paper manufacturing and
%wastewater management. Typically in these processes, mechanical dewatering techniques employing %pressure gradients across the suspension are favoured over thermal
%drying if possible, since the former methods are relatively rapid and energy efficient. However, %clogging of the dewatering filter and the lowered
%permeability of the suspended fibers, generally associated with the formation of a fibrous %network at the filter as dewatering proceeds,
%may significantly impede this process. Utilizing a continuum model for a general fiber %suspension calibrated against experimental work
%on commercial cellulose suspensions, we show here that switching from a constant pressure %difference profile to a temporal pulsing of the driving pressure difference
%provides a significant remedy to these problems.
%\pacs{47.57.Gc, 47.56.+r, 47.55.Kf}

% 47.57.Gc 	Granular flow
% 47.56.+r 	Flows through porous media
% 47.55.Kf 	Particle-laden flows

% 45.70.-n 	Granular systems
% 45.70.Mg 	Granular flow: mixing, segregation and stratification
%\keywords{dewatering, two-phase flow, experimental, numerical, cellulose, pressure pulsing}
%\maketitle
\end{frontmatter}

\cprotect\blfootnote{\textcopyright 2019. This manuscript version is made available under the CC-BY-NC-ND 4.0 license \verb|http://creativecommons.org/licenses/by-nc-nd/4.0/| }

\section{Introduction}
Numerous industrial fields, including wastewater treatment, paper manufacturing, and food industry employ
mechanical dewatering methods to process their respective products~\cite{neyens2003review,patist2008ultrasonic,alava2006physics}. While these products encompass a wide array of suspensions involving a variety of particle shapes and sizes, a particularly intriguing subset
among these is the class of fibrous suspensions, the rheological behavior of which can be governed by both colloidal and mechanical ({\it{e.g.}} frictious) aspects, depending on the fibre size and the chemical environment~\cite{schmid2000simulations}. Additional complications
may also be imposed by their geometrical anisotropy. Therefore, obtaining the optimal dewatering conditions for such suspensions is a challenge considering
the multitude of factors involved. However, some of these issues have been extensively studied in papermaking,
where it is crucial to control factors, such as the additives (chemicals) and fines content, affecting the mechanical dewatering of cellulose fiber suspensions in the forming section
of a paper machine~\cite{liimatainen2009influence,hubbe2007review,britt1985water,britt1986observations,britt1973mechanisms,britt1976new,britt1980water}.

%NFC paragraph:
Additionally, widespread substitution of petrochemical-based and other unsustainable materials in applications with biotic
ones requires higher functionality than accessible with traditional fiber-based products. Increasing
the functionality depends largely on the adaptation of novel advanced manufacturing paradigms.
An important route is by using nanotechnology, which allows fabrication of 
bioproducts from very small particles allowing for higher functionality.
Indeed over the recent years, the use of cellulose-based nanofibrillated materials have been one of the focus areas in bioengineering~\cite{kim2015}. These materials have large potential in several application fields, varying from their use as rheology modifiers \cite{BOLUK2011297} to reinforcing agents in nanobased composites \cite{kim2015,kargarzadeh2017recent}.
Further, a potential platform for large-scale production is reel-to-reel manufacturing similar to modern paper machines.
 Here the preparation process of composite structures involves dewatering. Unfortunately, the nanofiber-based suspensions have proven to be even more difficult to dewater than those based on macro-sized fibers~\cite{dimic2013role}. Thus, direct use of the forming methods familiar to paper production to generate the novel structures turns out to be unfeasible, and novel strategies must be researched.

To improve the efficiency of the dewatering processes associated with such fibrous suspensions, a theoretical model permitting numerical simulations is valuable.
It allows probing the response of the dewatered substance under varying conditions, which in turn would be either time consuming, expensive, or technically difficult to realize
in an experimental setup. 
Thus far, the numerical work on fiber suspensions seems to have revolved around the concept of Darcy's law~\cite{masoodi2010darcy,koponen2016flow}. In this continuum-level framework, the flow is assumed to be steady, neglecting dynamical effects, and the flow resistance of the fiber phase to the permeating liquid is mediated by suspension-specific permeability, a parameter that can depend on numerous quantities, such as fiber volume fraction, fiber-fiber interactions, particle shape, and filler content, for instance~\cite{koponen2016flow}. Accordingly, the bulk of the work performed in this framework has centered on forming empirical relationships for the permeability of fiber networks, treating the ensuing degrees of freedom of the model as fitting parameters~\cite{jackson1986permeability,koponen1998permeability,soltani2014effect}. Naturally, this prevents these models from being used {\it{a priori}} as experimental input is required.
Additionally, lattice Boltzmann methods have been successfully used in dealing with similar problems~\cite{guo2002lattice,lomine2013modeling}.
%% TODO: Näistä L-B jutuista lisää vielä
%% TODO: Tsekkaa: tässä varmasti kaikki olennaiset viitteet?

In this paper, we model a dewatering process involving an (initially) dilute fiber suspensions applying the full volume-averaged multiphase Navier-Stokes equations, establishing the flow resistance imposed by the forming fiber sheet via an effective drag correlation. We stress that the model is very flexible, allowing for dynamic effects, such as virtual mass forces, turbulence, or different drag correlations to be included readily, presently missing in the literature. Furthermore, the present approach permits us to simulate transient flows beyond the quasistatic limit as opposed to the earlier work, which will be important for simulating the pulsing phase, where the flow changes direction. Additionally, an analytical model independent of these numerical simulations is deduced from Darcy's law and modified to accommodate the pulsing events, as well.
Utilizing experimental data on MicroFibrillated Cellulose (MFC) and NanoFibrillated Cellulose (NFC), both the numerical and the analytical model are validated and displayed to be in excellent agreement with the real-world experiments.
Finally, the primary theme in this work is to demonstrate that temporal pulsing of the pressure difference driving the dewatering process provides both a superior averaged dewatering rate and final consistencies in the solid fiber sheet forming at the dewatering filter.

\section{Numerical model}
\begin{figure}[h]
\centering
 \includegraphics[width=0.4\textwidth]{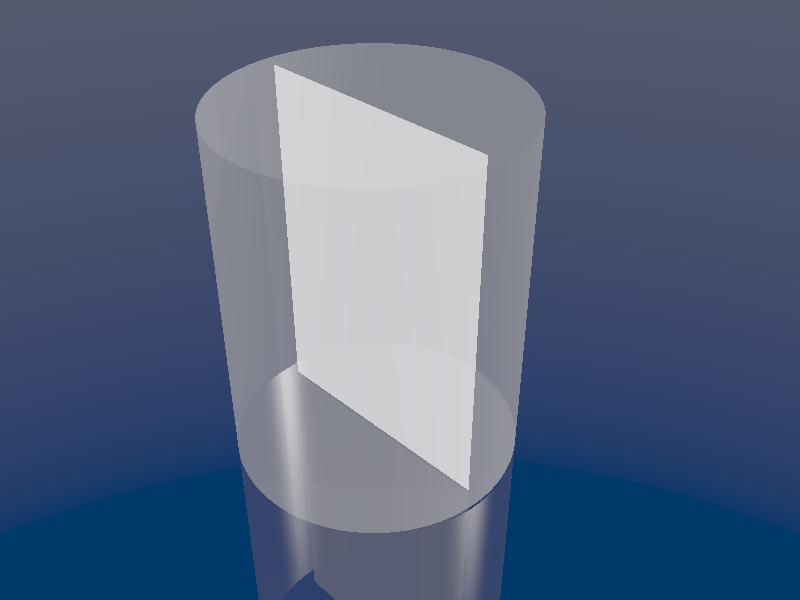}
 \caption{A schematic representation of the cross-section of a dewatering vessel utilized in the simulations (not to scale).}
 \label{fig:fig1extra}
\end{figure}
\begin{figure}[h]
\centering
 \includegraphics[width=0.4\textwidth]{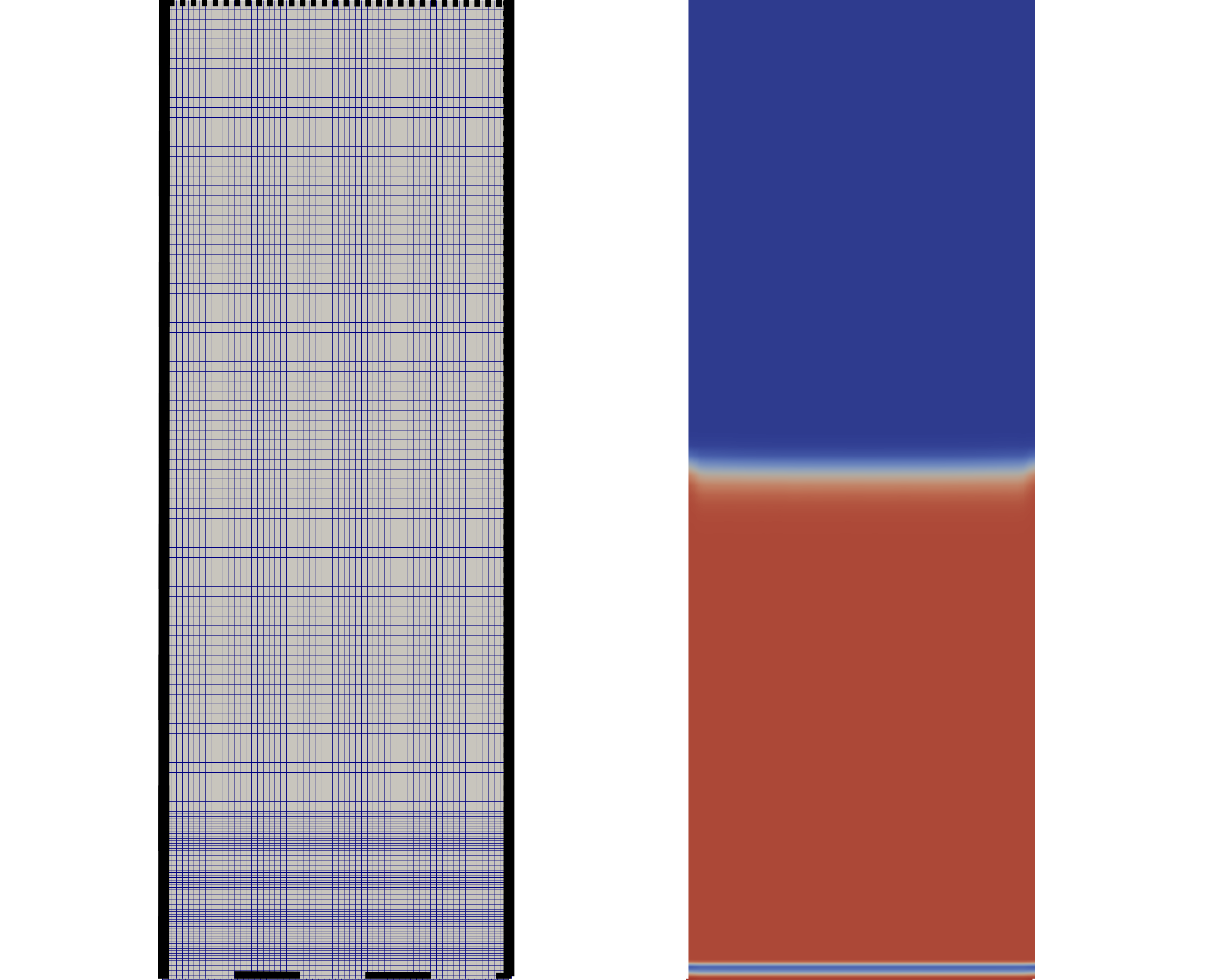}
 \caption{The mesh (left) of the cross-section presented in Fig.~\ref{fig:fig1extra}. The solid lines correspond to solid boundaries, the dashed line at the bottom to a semi-permeable boundary, and the slashed line on top to an air inlet. The figure on the right  displays the typical composition of phases inside the simulation geometry at the onset of dewatering: the red region consists of fibres and water while the blue region above it is occupied by air.}
 \label{fig:fig2extra}
\end{figure}
The Navier-Stokes equations provide a sound foundation for modeling fluids in the macroscopic continuum limit
and their volume-averaged forms, capable of accommodating the presence of multiple phases, were first introduced in depth by Anderson and Jackson~\cite{anderson1967fluid}. In essence, they treat the various phases as interpenetrating continuous media. These equations read
\begin{align}
    \frac{\partial \epsilon_i}{\partial t} + \nabla \cdot (\epsilon_i \mathbf{u}_i) &= 0 \label{eq:NS1}\\
    \rho_i\epsilon_i \left[\frac{\partial \mathbf{u}_i}{\partial t} + \mathbf{u}_i\nabla\mathbf{u}_i \right] &= -\nabla p + \nabla \cdot \mathbf{\tau} - \mathbf{f}_{ij} + \epsilon_i \rho_i \mathbf{g} ,
\label{eq:NS} \end{align}
where $\epsilon_i$ is the volume fraction of phase $i$, $\mathbf{u}_i$ is its respective velocity, $\rho_i$ its density, $p$ the pressure, $\tau$ the viscous stress tensor, $\mathbf{g}$ the gravity, and $\mathbf{f}_{ij}$ the interaction term with respect to a neighboring phase $j$. Together, these conservation of mass and momentum equations allow for evolving the system in time and solving the field variables detailing the motion of each phase.

As a typical dewatering process involves the interactions of air, liquid, and solid fibres embedded in that liquid, these three phases are modeled by Eqs.~\eqref{eq:NS1}~and~\eqref{eq:NS} with subscripts $a$, $f$ and $s$, respectively. For Eqs.~\eqref{eq:NS1}~and~\eqref{eq:NS} to provide unique numerical solutions for each phase, a constitutive relationship for $\tau$ is also required.
For this work, the general constitutive equation for the viscous stress $\tau = \eta_i (\nabla\mathbf{u} + \nabla\mathbf{u}^T )$ is sufficient. For each phase, a corresponding viscosity is defined, which is constant for the air and liquid phase. Importantly, the solid (fiber) phase is also modeled by the Navier-Stokes equations, and a ``viscosity'' for this phase is also provided. This viscosity should be perceived as the strength of the average viscous energy dissipation of the suspended phase.
However, since fibrous suspensions generally exhibit non-Newtonian behavior, presuming a constant value for this solid viscosity is hardly justified. Therefore, the dynamic viscosity for $\eta_s$ is modeled after the Krieger-Dougherty relation~\cite{krieger1959mechanism}
\begin{equation}
    \eta_s = \eta_0 \left(1 - \frac{\epsilon_s}{\epsilon_{max}} \right)^{-k} , \label{eq:KD}
\end{equation}
where $\eta_0$ and $k$ are fitting parameters
and $\epsilon_{max}$ refers to the jamming limit in terms of the solid (fiber) volume fraction.
% Remove the above sentence
 This relatively simple viscosity function provides the correct qualitative behavior of many fibrous complex fluids. 

The interaction term $\mathbf{f}_{ij}$ dictates the momentum exchange between phases via drag forces, virtual mass force, and lift forces, for instance. With regards to the dewatering flow, only drag $\mathbf{f}_{d}$ is assumed to contribute significantly to the momentum exchange. Furthermore, since this type of flow may involve rather high fiber concentrations, any mathematical drag correlation utilized here has to account for the presence ({\it{e.g.}}, drag screening) of neighboring particles, typically not included by simple Stokes drag. As such effects are difficult to include analytically, most drag correlations are empirically derived from experiments or micro-/mesoscale simulations. Accordingly~\cite{rusche2003computational}
\begin{align}
    \mathbf{f}_{d,ij} &= \frac{3}{4} \epsilon_i \frac{\rho_j}{d_i} C_d |\mathbf{u}_i-\mathbf{u}_j| \left(\mathbf{u}_i-\mathbf{u}_j\right) \label{eq:Drag} \\
    C_d &= \frac{24}{Re} \left(1 + 0.15 Re^{0.687} \right) , \label{eq:SN}
\end{align}
where the latter definition for the drag coefficient $C_d$ is suggested by Schiller and Naumann, %HH: Check: is the \textbf{definition} of the drag coefficient due to (=caused by) Schiller and Naumann??
who introduced this empirically obtained relation in 1933 by examining the drag imposed upon spherical fluid elements suspended in a surrounding carrier fluid. Here, $d_i$ is the characteristic hydrodynamic radius of the constituents in phase $i$ and the dimensionless particle Reynolds number $Re = \rho_j |\mathbf{u}_i-\mathbf{u}_j| d_i / \eta_j $. This correlation proves to be sufficiently accurate without posing excessively high computational costs.
In this framework, drag blending is also utilized, {\it{i.e.}}, the total drag experienced by phase ${\it{i}}$ is
\begin{equation}
    \mathbf{f}_{d,i} = \epsilon_j \mathrm{f}_{d,ij} + \epsilon_i \mathrm{f}_{d,ji} .
\end{equation}
The advantage of this blending is that the total drag remains both quantitatively and qualitatively sound even if a phase becomes the primary (suspending) or the secondary (suspended) fluid in a control volume in terms of the volume fraction $\epsilon_i$.

The geometry of a typical dewatering vessel is pictured in Fig.~\ref{fig:fig1extra}. In this work, the 2D simulations are performed on a geometry presenting the cross-section of an axisymmetric dewatering vessel seen in the figure. While this selection does not precisely match the experimental boundary conditions, we expect this choice to yield results qualitatively comparable with the experimental data. To match the typical suspension height in an experimental vessel, the length of the short sides in this cross section is chosen as $a = \sqrt{\pi r^2}$, where $r$ is the experimental vessel radius. The subsequent mesh and phase profile at the onset of the dewatering flow is presented in Fig.~\ref{fig:fig2extra}. The long sides are treated as solid walls with no-slip boundary conditions, while the upper side of the cross-section is a varying pressure air inlet. Additionally, the lower side is a constant pressure outlet, which allows water and air to permeate. However for the solid phase, a momentum sink is applied in the Navier-Stokes equations as soon as it enters the outlet, preventing the removal of the solid content via the outlet. Finally in the event of a pressure pulse, the lower side serves as a constant pressure inlet, and the upper side as a varying pressure outlet.

The simulations were performed on a mesh consisting of 400000 hexahedrons, and the mesh quality was increased in the vicinity of the wire, as seen in Fig.~\ref{fig:fig2extra}, to provide a better accuracy in examining the forming cake. The constant pressure term inside the vessel was set to atmospheric pressure and the pressure at the outlet was set to match the vacuum pressure produced by the vacuum pump in the experiments. In the pulsed simulations, a pulse time of 3 seconds was used, with the outlet pressure set to 2500 mbar during the pulse.

It is important to notice the essential difference of this approach in comparison to the more traditionally utilized Darcy or Darcy-Brinkman laws. The latter laws can also be understood as the volume-averaged expansion
of Navier-Stokes equations from the microscopic limit inside a porous medium element to the macroscopic limit, where the hydraulic conductivity $\kappa$ is then related to the average porosity of the medium~\cite{narsilio2009upscaling}. However, a possible issue associated with this framework is that $\kappa$ may be an ill-defined quantity for fibrous suspensions of a dilute concentration, where no permeable cross-linked structure (sheet) has yet formed, as is the case in the initial stages of a typical dewatering flow {\it{e.g.}}, in paper pulp dewatering in the forming section of a paper machine.

On the other hand, this average porosity can also be accommodated to a drag correlation function such as the one applied here by a correction factor $\Omega$ as noted in Ref.~\cite{wu1998hydrodynamic}, as the finite porosity of a solid object reduces the effective drag experienced by the otherwise solid structure. Therefore, the  approach presented here could complement the Darcy-like description, as one may also perceive the forming fiber structure (sheet) in a dewatering flow at the filter (wire) as a solid object with a finite porosity via the $\Omega$-factor. However, our approach is more minimalist still, as the Krieger-Dougherty expression~\eqref{eq:KD} for the effective viscosity of the fibrous suspension predicts rapidly increasing viscosity in the forming sheet, which is then translated to a rapidly growing drag/flow resistance in the Schiller-Naumann drag correlation~\eqref{eq:SN}.

% Limitations of the model
% Fibre water retention, transition to gel.
This numerical scheme also has some limitations in its capabilities to model drainage of interacting particle systems. Since the particle phase hydrodynamic volume fraction is conserved over the simulation, it neglects any aggregation-related structural properties that could lead to changes in the network water retention and eventual gelation \cite{puisto2012modeling}.

\section{Methods} \label{sec:sec2}
\subsection{Experimental setup}
\begin{figure}[h]
\centering
 \includegraphics[width=0.4\textwidth]{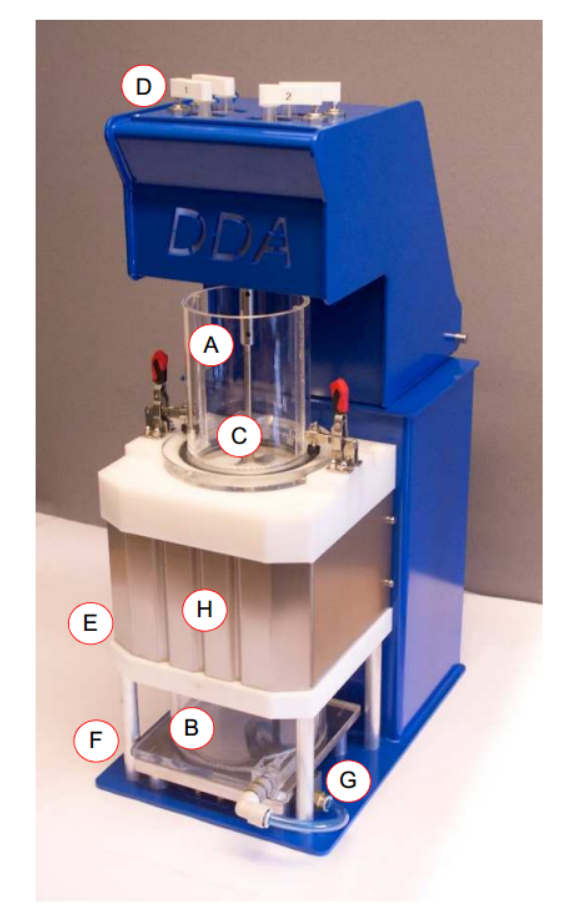}
 \caption{The dynamic drainage analyzer (DDA) utilized in producing the experimental data. The figure displays the dewatering vessel (A), vacuum vessel (B), stirrer (C), chemical dosage unit (D), flush water inlet (E), flush water outlet (F), turbidity cell (G), and the protective cover (H).}
 \label{fig:fig1}
\end{figure}
Although numerical in scope, the work presented here is supported by dewatering experiments performed in a Dynamic Drainage Analyzer (DDA), which closely reflects the actual dewatering conditions of industrial paper machines. As illustrated in Fig.~\ref{fig:fig1}, the primary components comprising the DDA are the reaction vessel, vacuum vessel, and the flush water inlet/outlet. The vacuum transmitter and pump (covered by the protective cover in Fig.~\ref{fig:fig1}) then provide the necessary pressure difference to drive the solvent from the suspension present in the reaction vessel, the bottom of which is augmented by a mesh of a size determined by the typical pore size of the suspension. Here a wire mesh of screen 0.025, thread size 0.025 mm and a mesh size of 500, was used. The setup and the dewatering experiments are controlled by software that allows adjusting the relevant parameters, such as the vacuum level and dewatering time, at will.

In this work, the MFC fibers (hardwood kraft pulp) were received from Suzano Pulp and Paper. The NFC fibers, prepared on-site, are once dried bleached birch kraft pulp, and the initial fiber slurry was soaked for one day and then dispersed using a high shear Diaf dissolver for 10 minutes at 700 rpm. Subsequently it was fed into Masuko Sangyo’s Supermasscolloider (Masuko Sangyo Co., Kawaguchi-city, Japan) type MKZA 10-15J, utilizing three passes.
\begin{table}[h]
    \centering
    \begin{tabular}{c|c|c} 
         Property & MFC & NFC  \\ \hline
         Shear viscosity [mPas] & 540.53 & 23173  \\
         Transmittance [\%] & 23.88 & 34.33 \\
         Fines content [\%] & 32.8 & -- \\
    \end{tabular}
    \caption{Relevant properties of the measured suspensions.}
    \label{tab:tab1}
\end{table}
Tab.~\ref{tab:tab1} summarizes the suspension properties relevant to this work. The shear viscosities of the suspensions were determined by the Brookfield viscometer (model DV2TRV, V-72 vane spindle for MFC and V-73 for NFC) at 1.5\% consistency and 10 rpm stirrer speed. The transmittance was obtained with a Shimadzu-UV-Vis spectophotometer 2550 ($\lambda = 800$ nm, at consistency 0.1 \%).

\subsection{Numerical schemes}
Utilizing the Finite Volume method (FVM) provided in the OpenFOAM\textregistered~\cite{weller1998tensorial} library, along with the pre-existing implementation of the multiphase Navier-Stokes equation and the drag correlations (Eqs. (1)-(2) and (4)-(5)), the model described in the previous section is implemented and the subsequent dewatering simulations performed in a geometry closely resembling the experimental reaction vessel presented in Fig.~\ref{fig:fig1}. Since the vessel exhibits cylindrical symmetry, the simulations are performed in a geometry matching the 2D cross-section of the vessel presented in Fig.~\ref{fig:fig1} to conserve computational resources, as explained in the previous section. This 2D projection contains the inlet area, the suspension region, the filter zone, and the outlet area (see Fig.~\ref{fig:fig2extra}). In a typical simulation, the simulation is initialized by filling the geometry with three distinct phases (air, water, fibers), each modeled by their individual set of Navier-Stokes equations (Eq.~\eqref{eq:NS}). Initially, the geometry is loaded with air excepting the suspension region, where liquid and fiber phases are included in a mixture devised according to the experimental suspensions. Then, a pressure difference is set between the inlet and outlet to induce flow of the suspension. Both air and water are allowed to flow via the filter zone, however, the fiber phase is prevented from permeating this zone, simulating the operation of the mesh wire in the DDA device.

Two distinct types of simulations are performed. In the first one, a constant pressure difference is maintained between the inlet and the outlet, therefore closely resembling the steady dewatering process obtained in the experimental DDA. However, the second simulation type employs a pulsing pressure scheme in which this steady pressure difference is maintained for a period of time and then inverted for a specified pulse time ($t_p$) to precipitate a temporary breakup of the forming fiber sheet at the filter zone and providing a drastically improved dewatering rate in the subsequent steady dewatering period.

\subsection{Analytical model}
\begin{figure}[tb]
\centering
 \includegraphics[width=0.38\textwidth]{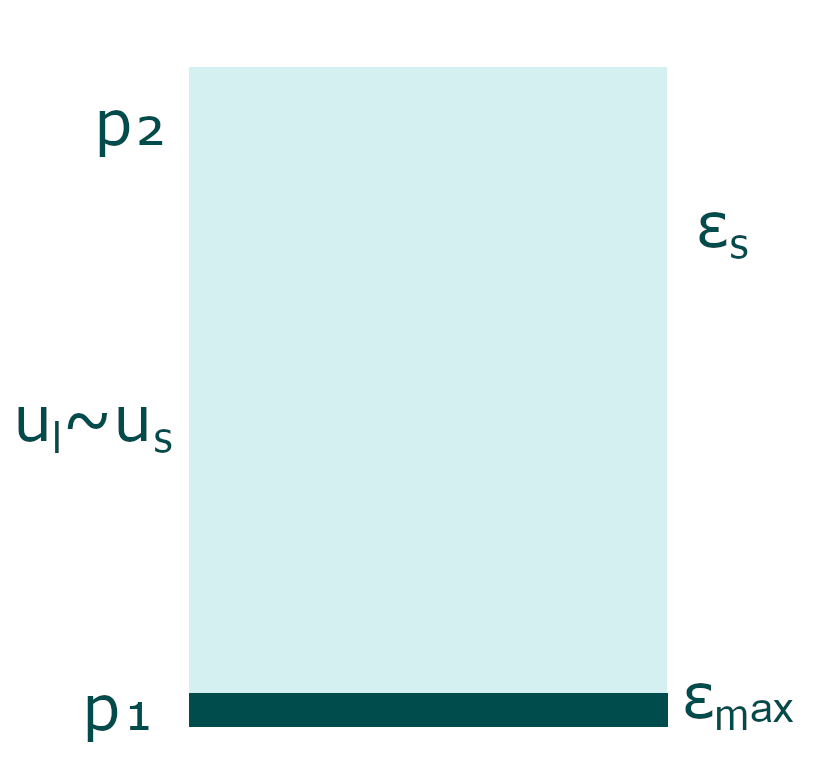}
 \caption{A schematic representation of a dewatering event in the context of the analytical model. A pressure difference $p_2-p_1$ exists between the inlet and the outlet, driving the water at a velocity $u_f$ towards the outlet. At steady-state, the same velocity is imposed on the solid content due to drag, $u_s \approx u_f$. The solid content (volume fraction $\epsilon_s$) therefore flows towards the wire, where it begins to form a cake with a volume fraction $\epsilon_{max}$. Importantly, it is assumed that during a steady dewatering event, or a steady dewatering event occuring between two successive pulses, $\epsilon_s$ is constant. An ideal pulse then disperses the solid content of the cake throughout the suspension, homogenizing the solid volume content. As a result, the solid content in the suspension region ($\epsilon_s$) varies only between two steady dewatering events that are separated by a pressure pulse.}
 \label{fig:figDarcy}
\end{figure}

It should be noted that the steady dewatering process can be viewed as a self-limiting Darcy flow. When the pressure difference is applied, the suspension flows towards the mesh wire and water exits, while an increasing number of fibers gather at the mesh wire at a concentration presumably corresponding to their maximum packing fraction ($\epsilon_{max}$) (see Fig.~\ref{fig:figDarcy}). As a result, the forming fiber sheet (cake) increases in height $H$ as the flow progresses in time $t$. Presented more precisely, the ensuing dewatering rate $q_f$ through the wire can be estimated by Darcy's law
\begin{align}
    q_f &= \frac{-k A \left(p_2 - p_1\right)}{\eta_s(\epsilon_{max}) H} =
    \frac{-k A \left(p_2 - p_1\right)}{\eta_s(\epsilon_{max}) \int_{t_0}^{t} \frac{\epsilon_s}{\epsilon_{max} A} q_f dt} = \nonumber \\
    &\frac{C}{\int_{t_0}^{t} \frac{\epsilon_s}{\epsilon_{max}} q_f dt} \label{eq:Darcy},
\end{align}
where $k$ is the permeability, $A$ is the cross-section area normal to the dewatering flow, $\eta_s(\epsilon_{max})$ is the viscosity of the suspension at the cake (provided by Eq.~\eqref{eq:KD}), and $p_2 - p_1$ is the (constant) pressure difference across the reaction vessel. Denoting the constant terms by $C$, one also observes that the integral describing the increase of $H$ can be expressed using the total fluid volume that has passed through the cake $Q_f$, that is, $H = \int_{t0}^{t} \left({\epsilon_s}/{\epsilon_{max}}\right) q_f dt = \left({\epsilon_s}/{\epsilon_{max}}\right) (Q_f(t) - Q_f(t_0)) = \left({\epsilon_s}/{\epsilon_{max}}\right) Q_f(t)$ (setting the arbitrary $t_0 = 0$) while clearly $q_f = \frac{\partial Q_f}{\partial t} = \dot{Q}_f(t)$. Moving this integrand to the left-hand side of Eq.~\eqref{eq:Darcy}, one then has
\begin{align}
    &\dot{Q}_f(t)Q_f(t) = \frac{1}{2} \frac{\partial}{\partial t} [Q_f(t)^2] = \frac{\epsilon_{max} C}{\epsilon_s} \nonumber \to \\ &Q_f(t) = \frac{2\sqrt{\epsilon_{max}}C \sqrt{t}}{\sqrt{\epsilon_s}} \label{eq:SLDarcyVolume}
\end{align}
and therefore (setting $D=2\sqrt{\epsilon_{max}}C$),
\begin{equation}
    \dot{Q}_f(t) = q_f = D \epsilon_s^{-1/2} t^{-1/2} . \label{eq:SLDarcy}
\end{equation}
Therefore, Eq.~\eqref{eq:SLDarcyVolume} implies a universal curve of the form $at^{1/2} + b$, expected to capture the steady dewatering flows.

Regarding the pressure pulsed dewatering flows, the outflow rate deduced above can be successfully utilized to yield an optimal number of pulsations in a single dewatering process. However, a number of additional conditions have to be imposed. These include
\begin{itemize}
\item After each pressure pulse, a fixed amount of fluid is removed before initiating the next pulse
\item This fixed amount of fluid removed between pulses is a small quantity in comparison to the initial suspension volume $V_0$ at $t=0$
\item Performing a single pulse takes a fixed amount of time $t_p$, which is a constant throughout the dewatering event.
\end{itemize}
Then it is relatively easy to deduce that the total dewatering time in a pulsed scheme is
\begin{equation} \label{eq:tottp}
    t_{tot,pulsed}=\left(\frac{1}{N} + \frac{F_{tot}}{2 N V_0} \right)T_{exp} + (N-1)t_p ,
\end{equation}
where $N-1 = n$, $n$ denoting the number of pulses, $F_{tot}$ is the total amount of fluid that is required to be removed, $V_0$ is the total volume of the suspension before commencing the dewatering process, and $T_{exp}$ is the typical experimental dewatering time in a constant pressure difference scheme without pulses. Now, this expression can be differentiated with respect to $N$ to yield the optimal number of pulsation steps $n$ in order to minimize the total dewatering time. That is, setting $\partial t_{tot,pulsed} / \partial N = 0$ will produce the optimum number of pressure pulses to dewater a suspension in as little time as possible. The full derivation of this is displayed in detail in Appx.~{A}.

\section{Results}

In the following figures, the computational results obtained with the framework discussed are presented. To obtain the best possible agreement with the experimental microfibrillated cellulose suspensions, the simulation suspension totals 1000 cm$^3$ of mixture containing 12\% fibers and 88\% water (in terms of volume fraction $\epsilon$). Matching exactly the particle phase volume fraction with that used to conduct the experiments requires detailed knowledge of the flexibility of the fibers, their aspect ratio, water-particle and particle-particle interactions, and the particle size distribution. In Ref.~\cite{puisto2012modeling}, a simple model has been used to condense this information into a single parameter, denoted as the effective mean aspect ratio of the cellulose fibers. Thus, matching the 12\% particle phase volume fraction used in the simulations to the initial mass concentration of the suspensions (0.034\%) used in the experiments would yield an effective aspect ratio of 27.5 in these simulations, which is a feasible figure for NFC fibers.
The density and viscosity of water and air are fixed to textbook values ($\rho_f = 1000$ kg/m$^3$, $\eta_f = 1$ mPas and $\rho_a = 1.2041$ kg/m$^3$, $\eta_a = 0.01813$ mPas, respectively) while the fiber density is $\rho_s = 1500$ kg/m$^3$ and $\eta_s(\epsilon_s=0) = \eta_0 = 10$ mPas.
The hydrodynamic radius for each phase (water, air, and particles) is set to $d_f = 1 \cdot 10^{-4}$ m, $d_a = 5 \cdot 10^{-5}$ m, and $d_s = 1 \cdot 10^{-4}$ m, respectively. As these should be treated as fitting parameters in the context of this work, these values were chosen to provide the best agreement with  experimental data. Furthermore, $\epsilon_{max}$ was also determined as a fitting parameter and set to a value of 0.9572 for best agreement with the experimental data. While this may seem excessively high, one must note that aside from their anisotropy, the size distribution of the fibers simulated here is very wide allowing them to pack extremely densely compared to, say, random sphere packings.
% Kommentoi vapaita parametrejä. aeff selostus.
\begin{figure}[h]
\centering
 \includegraphics[width=0.46\textwidth]{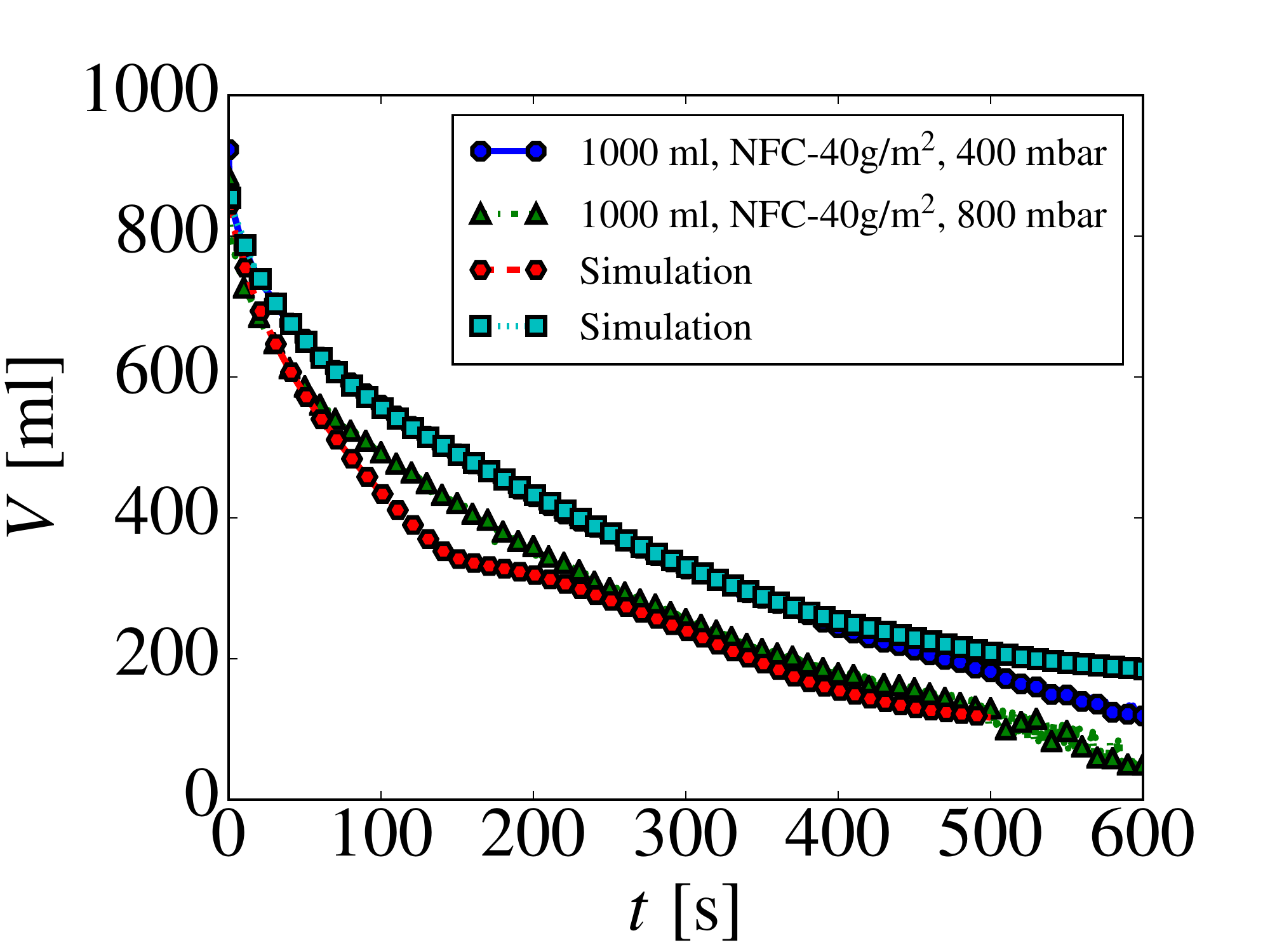}
 \caption{The remaining fluid volume $V$ inside the vessel as a function of the dewatering time $t$ in a constant pressure difference pulsing scheme. The simulations provide a delightful %HH: delightful!
 match with the experiments.}
 \label{fig:fig2}
\end{figure}

\begin{figure}[h]
\centering
 \includegraphics[width=0.46\textwidth]{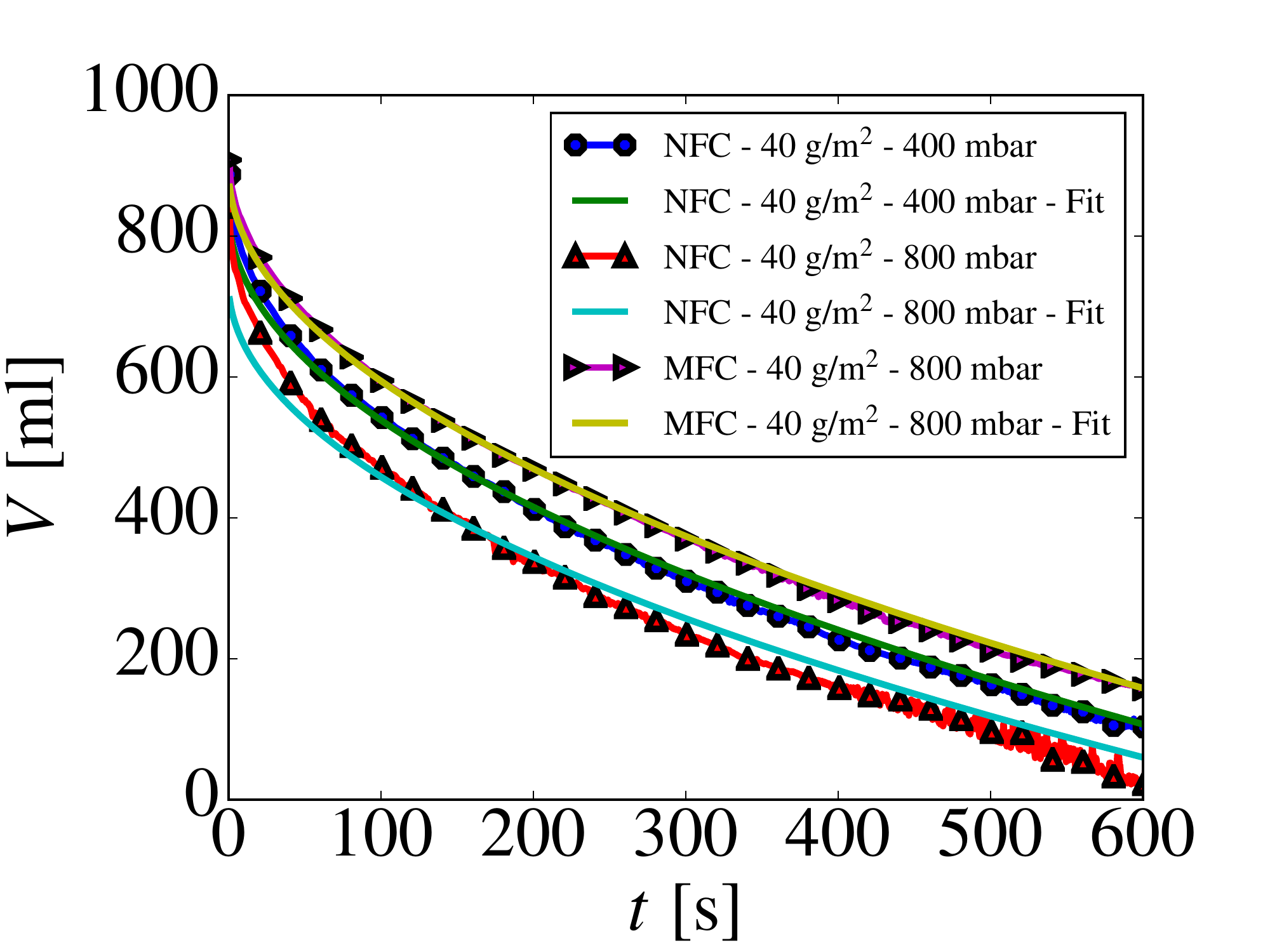}
 \caption{The remaining fluid volume $V$ inside the vessel as a function of the dewatering time $t$ in a constant pressure difference pulsing scheme. Judging by the figure, the analytical model (Eq.~\eqref{eq:SLDarcyVolume}) provides very good agreement with experimental data.}
 \label{fig:fig2-2}
\end{figure}
Examining the steady dewatering results first, Fig.~\ref{fig:fig2} displays the suspension volume $V$ present in the reaction vessel as a function of the dewatering time $t$ in both experiments and simulations. As the dewatering process proceeds with increasing $t$, the dewatering rate (the slope of the curves) decreases as the fiber sheet begins to form at the mesh wire, increasing the flow resistance towards the exiting water. Additionally, agreement with experiments and simulations seems excellent, lending credence to the computational approach applied in this work. Additionally in Fig.~\ref{fig:fig2-2}, the experimental data are also captured well by the universal curve of a self-limiting Darcy flow $f = a t^{1/2} + b$ presented in Sec.~\ref{sec:sec2} and fitted using the least squares method.

\begin{figure}[h]
\centering
 \includegraphics[width=0.46\textwidth]{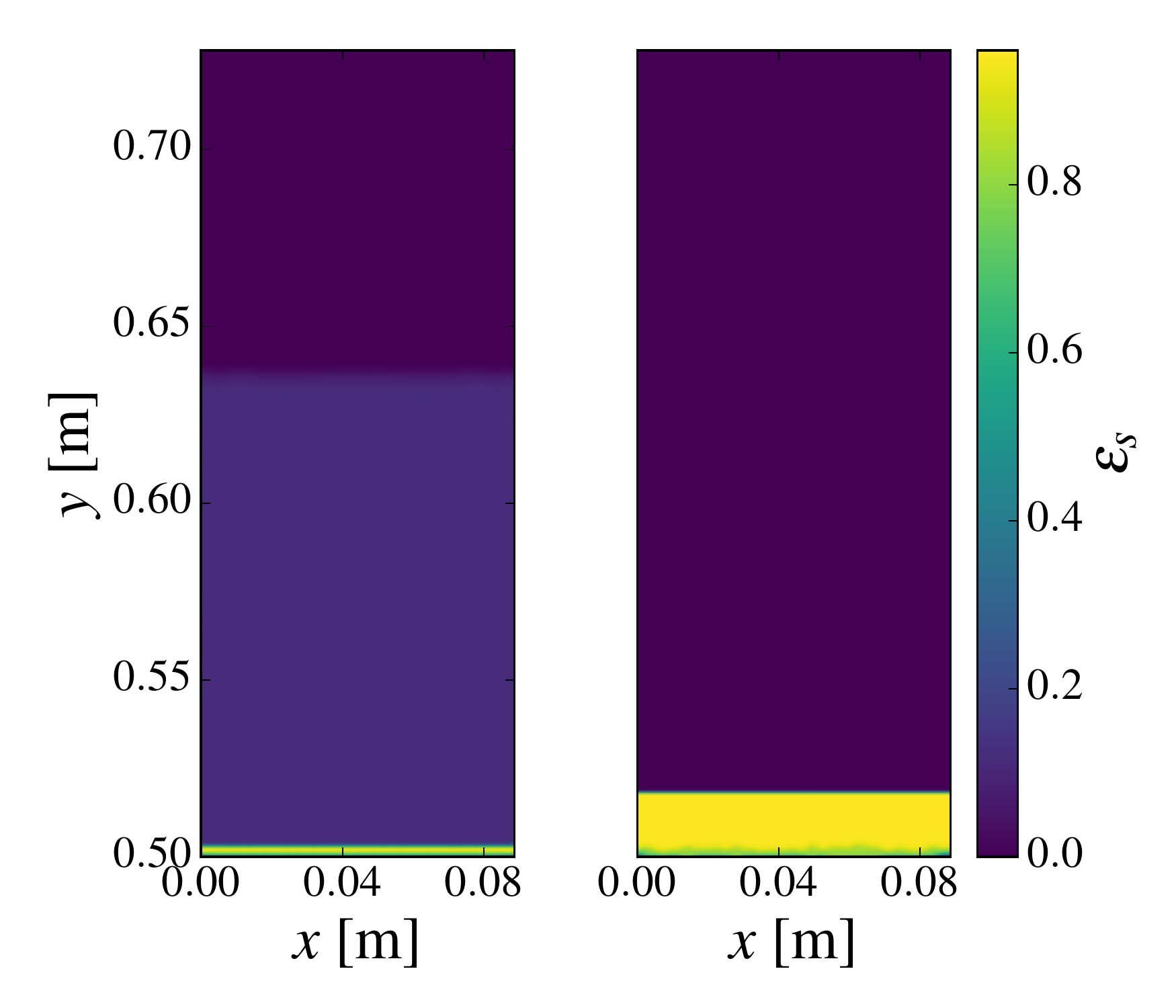}
 \caption{The cake formed in the steady pressure difference scheme. The color map depicts the volume fraction of the solid content on a $xy$-projection at the dewatering wire (located at $y=0.5$). On the left panel, around a second of dewatering has occurred. On the right panel, 200 seconds of dewatering flow has occurred.}
 \label{fig:fig3}
\end{figure}
Furthermore, Fig.~\ref{fig:fig3} displays the formation of the fiber sheet (cake) at the filter zone in the simulations at various occasions $t$ of such a steady dewatering process. The simulations show a uniform and gradual consolidation of a cake, which then diminishes the dewatering rate of the departing water.

\begin{figure}[h]
\centering
 \includegraphics[width=0.46\textwidth]{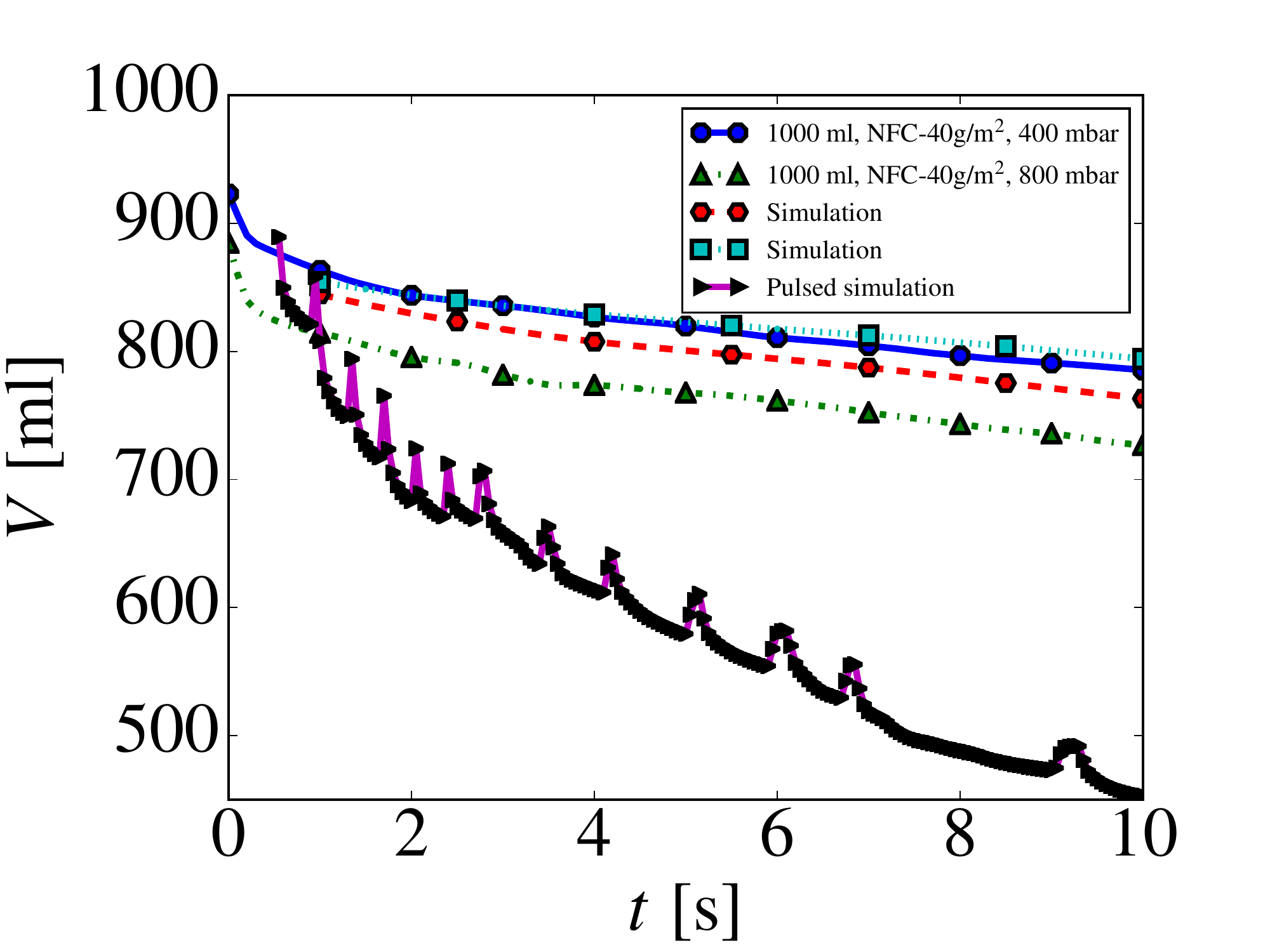}
 \caption{The effect of a typical pressure pulsing scheme on the dewatering process. As seen here, the pressure pulsing provides a superior dewatering rate to the steady pressure difference scenario.}
 \label{fig:fig4}
\end{figure}
Moving to the pulsed simulations, Fig.~\ref{fig:fig4} illustrates the suspension volume $V$ plotted as a function of the dewatering time $t$. Now however, a pulsed pressure scheme is included. As seen here, the pulsing scheme provides superior dewatering rates compared to the results obtained in Fig.~\ref{fig:fig2}. The improvement is an order of a magnitude decrease %HH: It is difficult to parse 'is on the order of a magnitude in terms of...'. Consider: 'The improvement is an order of magnitude better/faster/etc.'...
in terms of the dewatering time $t$, which should be considered a significant reduction over the typical dewatering times observed in a steady dewatering process.

\begin{figure}[h]
\centering
 \includegraphics[width=0.46\textwidth]{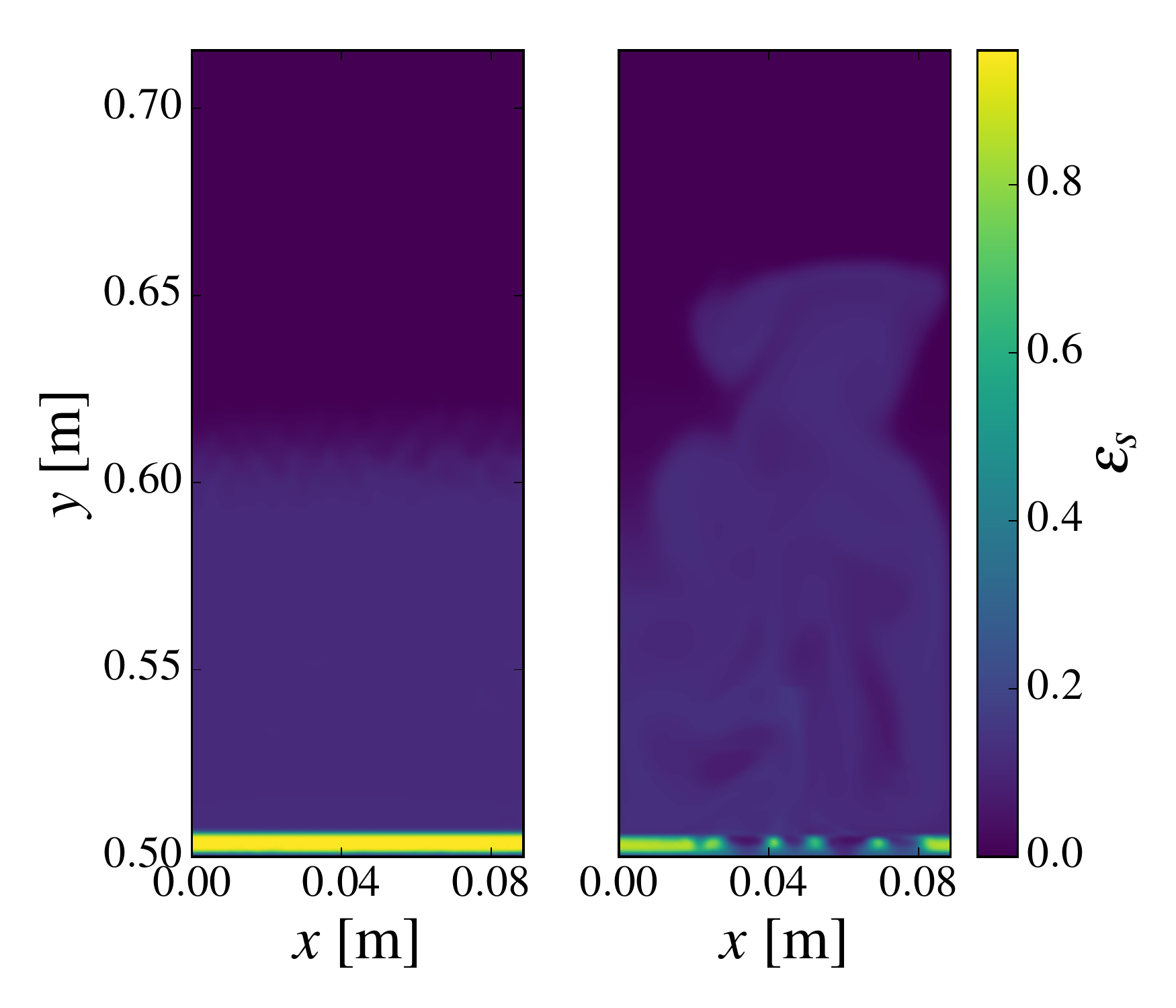}
 \caption{The cake formed in the pulsed pressure difference scheme. The color map depicts the volume fraction of the solid content on a $xy$-projection at the dewatering wire (located at $y=0.5$). On the left panel, around 10 seconds of dewatering has occurred. This corresponds to the exact time at which the pressure difference is zero ($p_2-p_1$ = 0) before a pulse ($p_1 > p_2$) is initiated. On the right side panel, 13.1 seconds of dewatering flow has occurred, corresponding to the time where the pressure pulse has redistributed the solid content and a steady dewatering is again taking place. The solid volume fraction forming the cake has disintegrated due to the pressure pulsing, and the flow resistance is reduced.}
 \label{fig:fig5}
\end{figure}
Furthermore, Fig.~\ref{fig:fig5} demonstrates the formation of the cake at the filter zone in a pulsed scheme in the simulations, where the subplots at various times are chosen to include a typical pulsing event. As seen here, the pulsing typically fragments the cake at random locations at the filter in general, and results in a more inhomogeneous cake structure in the proximity of the filter. However, these heterogeneities tend to equalize over time as a steady pressure difference is once again applied, implying the sheet structure is not compromised by the inclusion of the pressure pulsing mechanism.

\begin{figure}[h]
\centering
 \includegraphics[width=0.46\textwidth]{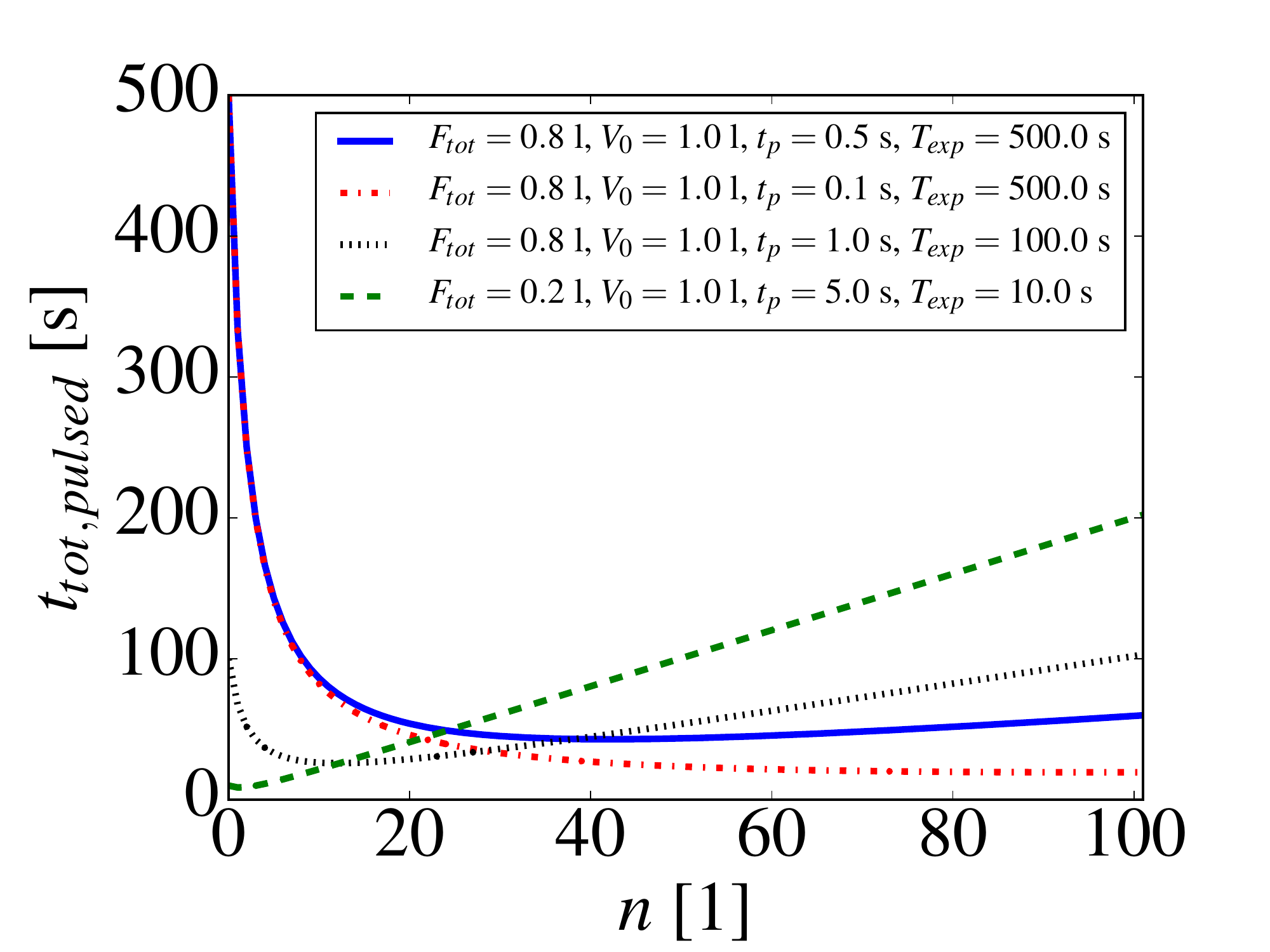}
 \caption{The total dewatering time in a pulsed scheme as expressed in Eq.~\eqref{eq:tottp}. As seen here, the total dewatering time  may or may not possess a minimum as a function of $n$ depending on the dewatering conditions, such as the water to be removed $F_{tot}$, initial suspensions volume $V_0$, and the time it takes to perform a single pulse $t_p$, as well as the typical steady dewatering total time $T_{exp}$. The figure was derived applying the sixth-order Taylor expansion of $\epsilon_s$ ($T_6(\epsilon_{s,i})$, see Appx. A).}
 \label{fig:fig6}
\end{figure}
Finally, Fig.~\ref{fig:fig6} represents the typical duration $t_{tot,pulsed}$ of a dewatering process in a pulsed dewatering scheme derived with the aid of the analytical model (see Eq.~\eqref{eq:tottp}). Increasing the number of pressure pulses initially decreases the time required to remove the prescribed amount of fluid ($F_{tot}$) from the suspension. However, increasing the number of pulses indefinitely eventually increases $t_{tot,pulsed}$ since the majority of the time consumed dewatering is then actually spent applying the pulses. Between these two extrema, the minimum value for $t_{tot,pulsed}$ is obtained with the optimized number of pulses $n$. In practice, applying this number of pulses will result in dewatering the suspension at hand in as little time as possible. Another striking feature of this figure is that whether $t_{tot,pulsed}$ exhibits a unique minimum as a function of the pulse steps $n$ depends on the exact dewatering conditions. In general, these are set in the model by the initial volume of the suspension $V_0$, the water volume to be removed $F_{tot}$, and the time it takes for the dewatering device to perform a single pulsing $t_p$ as well as a typical steady dewatering event $T_{exp}$. Therefore, a pulsing scheme may or may not optimize the dewatering process depending on the dewatering device and suspension characteristics.
%Strikingly, secondary flows seems to be present at the proximity of the filter at this pulsing step, possibly %hinting at turbulent flow conditions during the pulsing event. We stress that such events are difficult to %capture with a simple Darcy-type expression, and necessitate the use of the full Navier-Stokes equations.

\section{Conclusions}
Dewatering processes are ubiquitous in various industrial settings. In mechanical dewatering utilizing a pressure driven dewatering flow, great care must be exercised while optimizing these flows to ensure a high dewatering rate while simultaneously attending to process-specific requirements, such as high and uniform sheet consistency in the paper-making industry, for instance. Here, supported by experimental dewatering results of nanofibrillated cellulose (NFC) obtained in a Dynamic Drainage Analyzer (DDA), it is shown that a continuum level description using volume-averaged multiphase Navier-Stokes equations enchanced with general drag correlations capture the essential dewatering characteristics extremely well. The model also allows dynamic effects, such as virtual mass forces and turbulence, to be accounted for with ease.

Additionally, we demonstrate that enhancing the steady dewatering process by a simple, periodic inversion of the driving pressure, denoted as pressure pulsing, decreases the required dewatering time by an order of magnitude for a typical NFC suspension while maintaining a uniform and high sheet consistency. The simulations also imply the possibility of turbulent flow conditions inside the dewatering vessel during a pulsing event. 

\section{Acknowledgements}
We wish to thank the Jane and Aatos Erkko Foundation for their financial support via the NANOFORM project and the Aalto Science-IT project for the high-performance computational resources. The financial support of the Academy of Finland (Project No. 278367) is also greatly appreciated.

%\addcontentsline{toc}{section}{References}
\bibliographystyle{elsarticle-num-names} 
\bibliography{master}

\begin{thebibliography}{29}
\providecommand{\natexlab}[1]{#1}
\providecommand{\url}[1]{\texttt{#1}}
\providecommand{\urlprefix}{URL }
\expandafter\ifx\csname urlstyle\endcsname\relax
  \providecommand{\doi}[1]{doi:\discretionary{}{}{}#1}\else
  \providecommand{\doi}[1]{doi:\discretionary{}{}{}\begingroup
  \urlstyle{rm}\url{#1}\endgroup}\fi
\providecommand{\bibinfo}[2]{#2}

\bibitem[{Neyens and Baeyens(2003)}]{neyens2003review}
\bibinfo{author}{E.~Neyens}, \bibinfo{author}{J.~Baeyens}, \bibinfo{title}{A
  review of thermal sludge pre-treatment processes to improve dewaterability},
  \bibinfo{journal}{J. Hazard. Mater.}
  \bibinfo{volume}{98}~(\bibinfo{number}{1-3}) (\bibinfo{year}{2003})
  \bibinfo{pages}{51--67}.

\bibitem[{Patist and Bates(2008)}]{patist2008ultrasonic}
\bibinfo{author}{A.~Patist}, \bibinfo{author}{D.~Bates},
  \bibinfo{title}{Ultrasonic innovations in the food industry: From the
  laboratory to commercial production}, \bibinfo{journal}{Innov. Food. Sci.
  Emerg. Technol.} \bibinfo{volume}{9}~(\bibinfo{number}{2})
  (\bibinfo{year}{2008}) \bibinfo{pages}{147--154}.

\bibitem[{Alava and Niskanen(2006)}]{alava2006physics}
\bibinfo{author}{M.~Alava}, \bibinfo{author}{K.~Niskanen}, \bibinfo{title}{The
  physics of paper}, \bibinfo{journal}{Rep. Prog. Phys.}
  \bibinfo{volume}{69}~(\bibinfo{number}{3}) (\bibinfo{year}{2006})
  \bibinfo{pages}{669}.

\bibitem[{Schmid et~al.(2000)Schmid, Switzer, and
  Klingenberg}]{schmid2000simulations}
\bibinfo{author}{C.~Schmid}, \bibinfo{author}{L.~Switzer},
  \bibinfo{author}{D.~Klingenberg}, \bibinfo{title}{Simulations of fiber
  flocculation: Effects of fiber properties and interfiber friction},
  \bibinfo{journal}{J. Rheol.} \bibinfo{volume}{44}~(\bibinfo{number}{4})
  (\bibinfo{year}{2000}) \bibinfo{pages}{781--809}.

\bibitem[{Liimatainen et~al.(2009)Liimatainen, Haavisto, Haapala, and
  Niinim{\"a}ki}]{liimatainen2009influence}
\bibinfo{author}{H.~Liimatainen}, \bibinfo{author}{S.~Haavisto},
  \bibinfo{author}{A.~Haapala}, \bibinfo{author}{J.~Niinim{\"a}ki},
  \bibinfo{title}{Influence of adsorbed and dissolved carboxymethyl cellulose
  on fibre suspension dispersing, dewaterability, and fines retention},
  \bibinfo{journal}{Bioresources} \bibinfo{volume}{4}~(\bibinfo{number}{1})
  (\bibinfo{year}{2009}) \bibinfo{pages}{321--340}.

\bibitem[{Hubbe and Heitmann(2007)}]{hubbe2007review}
\bibinfo{author}{M.~Hubbe}, \bibinfo{author}{J.~Heitmann},
  \bibinfo{title}{Review of factors affecting the release of water from
  cellulosic fibers during paper manufacture}, \bibinfo{journal}{Bioresources}
  \bibinfo{volume}{2}~(\bibinfo{number}{3}) (\bibinfo{year}{2007})
  \bibinfo{pages}{500--533}.

\bibitem[{Britt and Unbehend(1985)}]{britt1985water}
\bibinfo{author}{K.~Britt}, \bibinfo{author}{J.~Unbehend},
  \bibinfo{title}{Water removal during paper formation},
  \bibinfo{journal}{Tappi J.} \bibinfo{volume}{68}~(\bibinfo{number}{4})
  (\bibinfo{year}{1985}) \bibinfo{pages}{104--107}.

\bibitem[{Britt et~al.(1986)Britt, Unbehend, and
  Shridharan}]{britt1986observations}
\bibinfo{author}{K.~Britt}, \bibinfo{author}{J.~Unbehend},
  \bibinfo{author}{R.~Shridharan}, \bibinfo{title}{Observations on water
  removal in papermaking}, \bibinfo{journal}{Tappi J.}
  \bibinfo{volume}{69}~(\bibinfo{number}{7}) (\bibinfo{year}{1986})
  \bibinfo{pages}{76--79}.

\bibitem[{Britt(1973)}]{britt1973mechanisms}
\bibinfo{author}{K.~Britt}, \bibinfo{title}{Mechanisms of retention during
  paper formation}, \bibinfo{journal}{Tappi J.}
  \bibinfo{volume}{56}~(\bibinfo{number}{10}) (\bibinfo{year}{1973})
  \bibinfo{pages}{46--50}.

\bibitem[{Britt and Unbehend(1976)}]{britt1976new}
\bibinfo{author}{K.~Britt}, \bibinfo{author}{J.~Unbehend}, \bibinfo{title}{New
  methods for monitoring retention [of fines and filler, paper machines].},
  \bibinfo{journal}{Tappi J.} .

\bibitem[{Britt and Unbehend(1980)}]{britt1980water}
\bibinfo{author}{K.~Britt}, \bibinfo{author}{J.~Unbehend},
  \bibinfo{title}{Water removal during sheet formation [Paper making
  process].}, \bibinfo{journal}{Journal of the Technical Association of the
  Pulp and Paper Industry} .

\bibitem[{Kim et~al.(2015)Kim, Shim, Kim, Lee, Min, Jang, Abas, and
  Kim}]{kim2015}
\bibinfo{author}{J.-H. Kim}, \bibinfo{author}{B.~S. Shim},
  \bibinfo{author}{H.~S. Kim}, \bibinfo{author}{Y.-J. Lee},
  \bibinfo{author}{S.-K. Min}, \bibinfo{author}{D.~Jang},
  \bibinfo{author}{Z.~Abas}, \bibinfo{author}{J.~Kim}, \bibinfo{title}{Review
  of nanocellulose for sustainable future materials},
  \bibinfo{journal}{International Journal of Precision Engineering and
  Manufacturing-Green Technology} \bibinfo{volume}{2}~(\bibinfo{number}{2})
  (\bibinfo{year}{2015}) \bibinfo{pages}{197--213}.

\bibitem[{Boluk et~al.(2011)Boluk, Lahiji, Zhao, and McDermott}]{BOLUK2011297}
\bibinfo{author}{Y.~Boluk}, \bibinfo{author}{R.~Lahiji},
  \bibinfo{author}{L.~Zhao}, \bibinfo{author}{M.~T. McDermott},
  \bibinfo{title}{Suspension viscosities and shape parameter of cellulose
  nanocrystals (CNC)}, \bibinfo{journal}{Colloids and Surfaces A:
  Physicochemical and Engineering Aspects}
  \bibinfo{volume}{377}~(\bibinfo{number}{1}) (\bibinfo{year}{2011})
  \bibinfo{pages}{297 -- 303}.

\bibitem[{Kargarzadeh et~al.(2017)Kargarzadeh, Mariano, Huang, Lin, Ahmad,
  Dufresne, and Thomas}]{kargarzadeh2017recent}
\bibinfo{author}{H.~Kargarzadeh}, \bibinfo{author}{M.~Mariano},
  \bibinfo{author}{J.~Huang}, \bibinfo{author}{N.~Lin},
  \bibinfo{author}{I.~Ahmad}, \bibinfo{author}{A.~Dufresne},
  \bibinfo{author}{S.~Thomas}, \bibinfo{title}{Recent developments on
  nanocellulose reinforced polymer nanocomposites: A review},
  \bibinfo{journal}{Polymer} \bibinfo{volume}{132} (\bibinfo{year}{2017})
  \bibinfo{pages}{368--393}.

\bibitem[{Dimic-Misic et~al.(2013)Dimic-Misic, Puisto, Gane, Nieminen, Alava,
  Paltakari, and Maloney}]{dimic2013role}
\bibinfo{author}{K.~Dimic-Misic}, \bibinfo{author}{A.~Puisto},
  \bibinfo{author}{P.~Gane}, \bibinfo{author}{K.~Nieminen},
  \bibinfo{author}{M.~Alava}, \bibinfo{author}{J.~Paltakari},
  \bibinfo{author}{T.~Maloney}, \bibinfo{title}{The role of MFC/NFC swelling in
  the rheological behavior and dewatering of high consistency furnishes},
  \bibinfo{journal}{Cellulose} \bibinfo{volume}{20}~(\bibinfo{number}{6})
  (\bibinfo{year}{2013}) \bibinfo{pages}{2847--2861}.

\bibitem[{Masoodi and Pillai(2010)}]{masoodi2010darcy}
\bibinfo{author}{R.~Masoodi}, \bibinfo{author}{K.~Pillai},
  \bibinfo{title}{Darcy's law-based model for wicking in paper-like swelling
  porous media}, \bibinfo{journal}{AIChE J.}
  \bibinfo{volume}{56}~(\bibinfo{number}{9}) (\bibinfo{year}{2010})
  \bibinfo{pages}{2257--2267}.

\bibitem[{Koponen et~al.(2016)Koponen, Haavisto, Liukkonen, and
  Salmela}]{koponen2016flow}
\bibinfo{author}{A.~Koponen}, \bibinfo{author}{S.~Haavisto},
  \bibinfo{author}{J.~Liukkonen}, \bibinfo{author}{J.~Salmela},
  \bibinfo{title}{The flow resistance of fiber sheet during initial
  dewatering}, \bibinfo{journal}{Drying Technol.}
  \bibinfo{volume}{34}~(\bibinfo{number}{12}) (\bibinfo{year}{2016})
  \bibinfo{pages}{1521--1533}.

\bibitem[{Jackson and James(1986)}]{jackson1986permeability}
\bibinfo{author}{G.~Jackson}, \bibinfo{author}{D.~James}, \bibinfo{title}{The
  permeability of fibrous porous media}, \bibinfo{journal}{Can. J. Chem. Eng.}
  \bibinfo{volume}{64}~(\bibinfo{number}{3}) (\bibinfo{year}{1986})
  \bibinfo{pages}{364--374}.

\bibitem[{Koponen et~al.(1998)Koponen, Kandhai, Hellen, Alava, Hoekstra,
  Kataja, Niskanen, Sloot, and Timonen}]{koponen1998permeability}
\bibinfo{author}{A.~Koponen}, \bibinfo{author}{D.~Kandhai},
  \bibinfo{author}{E.~Hellen}, \bibinfo{author}{M.~Alava},
  \bibinfo{author}{A.~Hoekstra}, \bibinfo{author}{M.~Kataja},
  \bibinfo{author}{K.~Niskanen}, \bibinfo{author}{P.~Sloot},
  \bibinfo{author}{J.~Timonen}, \bibinfo{title}{Permeability of
  three-dimensional random fiber webs}, \bibinfo{journal}{Phys. Rev. Lett.}
  \bibinfo{volume}{80}~(\bibinfo{number}{4}) (\bibinfo{year}{1998})
  \bibinfo{pages}{716}.

\bibitem[{Soltani et~al.(2014)Soltani, Johari, and
  Zarrebini}]{soltani2014effect}
\bibinfo{author}{P.~Soltani}, \bibinfo{author}{M.~Johari},
  \bibinfo{author}{M.~Zarrebini}, \bibinfo{title}{Effect of 3D fiber
  orientation on permeability of realistic fibrous porous networks},
  \bibinfo{journal}{Powder Technol.} \bibinfo{volume}{254}
  (\bibinfo{year}{2014}) \bibinfo{pages}{44--56}.

\bibitem[{Guo and Zhao(2002)}]{guo2002lattice}
\bibinfo{author}{Z.~Guo}, \bibinfo{author}{T.~Zhao}, \bibinfo{title}{Lattice
  Boltzmann model for incompressible flows through porous media},
  \bibinfo{journal}{Phys. Rev. E} \bibinfo{volume}{66}~(\bibinfo{number}{3})
  (\bibinfo{year}{2002}) \bibinfo{pages}{036304}.

\bibitem[{Lomin{\'e} et~al.(2013)Lomin{\'e}, Scholtes, Sibille, and
  Poullain}]{lomine2013modeling}
\bibinfo{author}{F.~Lomin{\'e}}, \bibinfo{author}{L.~Scholtes},
  \bibinfo{author}{L.~Sibille}, \bibinfo{author}{P.~Poullain},
  \bibinfo{title}{Modeling of fluid--solid interaction in granular media with
  coupled lattice Boltzmann/discrete element methods: application to piping
  erosion}, \bibinfo{journal}{Int. J. Numer. Anal. Methods Geomech.}
  \bibinfo{volume}{37}~(\bibinfo{number}{6}) (\bibinfo{year}{2013})
  \bibinfo{pages}{577--596}.

\bibitem[{Anderson and Jackson(1967)}]{anderson1967fluid}
\bibinfo{author}{T.~Anderson}, \bibinfo{author}{R.~Jackson},
  \bibinfo{title}{Fluid mechanical description of fluidized beds. {E}quations
  of motion}, \bibinfo{journal}{Ind. Eng. Chem. Fundam.}
  \bibinfo{volume}{6}~(\bibinfo{number}{4}) (\bibinfo{year}{1967})
  \bibinfo{pages}{527--539}.

\bibitem[{Krieger(1959)}]{krieger1959mechanism}
\bibinfo{author}{T.~Krieger, I.and~Dougherty}, \bibinfo{title}{A mechanism for
  non--{N}ewtonian flow in suspensions of rigid spheres},
  \bibinfo{journal}{Trans. Soc. Rheol.}
  \bibinfo{volume}{3}~(\bibinfo{number}{1}) (\bibinfo{year}{1959})
  \bibinfo{pages}{137--152}.

\bibitem[{Rusche(2003)}]{rusche2003computational}
\bibinfo{author}{H.~Rusche}, \bibinfo{title}{Computational fluid dynamics of
  dispersed two-phase flows at high phase fractions}, Ph.D. thesis,
  \bibinfo{school}{Imperial College London (University of London)},
  \bibinfo{year}{2003}.

\bibitem[{Narsilio et~al.(2009)Narsilio, Buzzi, Fityus, Yun, and
  Smith}]{narsilio2009upscaling}
\bibinfo{author}{G.~Narsilio}, \bibinfo{author}{O.~Buzzi},
  \bibinfo{author}{S.~Fityus}, \bibinfo{author}{T.~Yun},
  \bibinfo{author}{D.~Smith}, \bibinfo{title}{Upscaling of {N}avier--{S}tokes
  equations in porous media: theoretical, numerical and experimental approach},
  \bibinfo{journal}{Comput. Geotech.}
  \bibinfo{volume}{36}~(\bibinfo{number}{7}) (\bibinfo{year}{2009})
  \bibinfo{pages}{1200--1206}.

\bibitem[{Wu and Lee(1998)}]{wu1998hydrodynamic}
\bibinfo{author}{R.~Wu}, \bibinfo{author}{D.~Lee}, \bibinfo{title}{Hydrodynamic
  drag force exerted on a moving floc and its implication to free-settling
  tests}, \bibinfo{journal}{Water Res.}
  \bibinfo{volume}{32}~(\bibinfo{number}{3}) (\bibinfo{year}{1998})
  \bibinfo{pages}{760--768}.

\bibitem[{Puisto et~al.(2012)Puisto, Illa, Mohtaschemi, and
  Alava}]{puisto2012modeling}
\bibinfo{author}{A.~Puisto}, \bibinfo{author}{X.~Illa},
  \bibinfo{author}{M.~Mohtaschemi}, \bibinfo{author}{M.~Alava},
  \bibinfo{title}{Modeling the rheology of nanocellulose suspensions},
  \bibinfo{journal}{Nordic Pulp \& Paper Research Journal}
  \bibinfo{volume}{27}~(\bibinfo{number}{2}) (\bibinfo{year}{2012})
  \bibinfo{pages}{277--281}.

\bibitem[{Weller et~al.(1998)Weller, Tabor, Jasak, and
  Fureby}]{weller1998tensorial}
\bibinfo{author}{H.~Weller}, \bibinfo{author}{G.~Tabor},
  \bibinfo{author}{H.~Jasak}, \bibinfo{author}{C.~Fureby}, \bibinfo{title}{A
  tensorial approach to computational continuum mechanics using object-oriented
  techniques}, \bibinfo{journal}{Comput. Phys.}
  \bibinfo{volume}{12}~(\bibinfo{number}{6}) (\bibinfo{year}{1998})
  \bibinfo{pages}{620--631}.

\end{thebibliography}

\newpage

\appendix
\section{} \label{sec:AppA}
The result in Eq.~\eqref{eq:tottp} can be derived by first noting that the homogenized volume fraction of particle at a any given time is
\begin{equation}
    \epsilon_s =\epsilon_{s,0} \frac{1}{1 - F_{tot}/V_0} ,
\end{equation}
where $F_{tot}$ is the fluid that has been removed up to that time, $\epsilon_{s,0}$ denotes the (homogeneous) volume fraction of the solid phase before any fluid is removed, and $V_0$ is the initial suspension volume. This definition corresponds to the average (bulk) solid volume fraction, if the remaining solid content in the suspension is dispersed evenly among the remaining suspension volume (which is what occurs in an idealized pressure pulsing event). Indeed, we assume that primary source of the flow resistance in a typical dewatering process is the dense cake forming at the wire, and an ideal pressure pulse redistributes this solid content forming the cake evenly among the suspension volume, therefore decreasing the resistance to dewatering. The subsequent removal of water is then significantly eased until another cake forms, necessitating another pressure pulse.  Now, we may assume that the removed water is fixed to portions of $F_{tot}/N$ (between successive pulses, a fixed amount of water is removed). If a number of $n$ pulses is applied, then $N = n+1$, and the total fluid removed at $t_i$ is clearly $\left[(i-1)/N\right] \cdot F_{tot}$. Therefore, between successive pulses, the bulk solid volume fraction is
\begin{equation}
     \epsilon_{s,i} = \epsilon_{s,0} \frac{1}{1 - \left[ (i-1) F_{tot} \right] / \left[ N V_0 \right]} ,
\end{equation}
which has the form $1 / \left(1 - x \right)$. This can expanded in a Taylor (power) series of the form $1 + x + x^2 + x^3 + ...$, where $x = \left[(i-1)F_{tot}\right] / NV_0$. We denote this series as $T_q(\epsilon_{s,i})$, where $q$ refers to the order of the expansion. 
Now, using Eq.~\eqref{eq:SLDarcyVolume} and expanding the volume fraction term in the first order ($T_1(\epsilon_{s,i})$), one has
\begin{align}
    \sum_{i=1}^{N} t_i = & \frac{\epsilon_{s,0}}{\epsilon_{max}(2C)^2} \sum_{i=1}^{N} \left(1 + \frac{(i-1) F_{tot}}{N V_0}\right) \frac{F_{tot}^2}{N^2} = \nonumber\\
    & \frac{\epsilon_{s,0} F_{tot}^2}{\epsilon_{max}(2C)^2} \left( \frac{1}{N} + \frac{N(N-1) F_{tot}}{2 N^3 V_0} \right) \approx \nonumber \\ 
    &\frac{\epsilon_{s,0} F_{tot}^2}{\epsilon_{max}(2C)^2} \left( \frac{1}{N} + \frac{F_{tot}}{2 V_0 N} \right) ,
\end{align}
where the first-order Taylor expansion of $\epsilon_s$ is summed as an arithmetic series. If higher-order terms were included, Faulhaber's formula could be applied to evaluate the power series sums emerging from these terms.
Now, dividing this by the theoretical dewatering time $t_0$ of a steady pressure difference scheme without pulses (obtained with Eq.~\eqref{eq:SLDarcyVolume}) and multiplying by the experimental steady dewatering time $T_{exp}$
\begin{align}
    &\left(\frac{\sum_{i=1}^{N} t_i}{t_0}\right) T_{exp} = \nonumber \\
    &\left(\frac{\epsilon_{s,0} F_{tot}^2}{\epsilon_{max}(2C)^2} \left( \frac{1}{N} + \frac{F_{tot}}{2 V_0 N} \right) / \left( \frac{\epsilon_{s,0}}{\epsilon_{max}(2C)^2}F_{tot}^2 \right)\right) T_{exp} = \nonumber \\
    & \left(\frac{1}{N} + \frac{F_{tot}}{2 V_0 N} \right) T_{exp} .
\end{align}
Finally, the total dewatering duration in a pulsed scheme is obtained by adding the finite (constant) time $t_p$ it takes to for the device to perform a single pulse multiplied by the number of pulses ($n=N-1$) to the expression above
\begin{equation}
    t_{tot,pulsed} = \left(\frac{1}{N} + \frac{F_{tot}}{2 V_0 N} \right)T_{exp} + (N-1)t_p ,
\end{equation}
which is then the total dewatering time derived with the first-order Taylor expansion $T_1(\epsilon_{s,i})$. Seeking the minimum of this function by setting $\partial t_{tot,pulsed} / \partial N = 0$ recovers the optimum number of pressure pulses that should be used to dewater a suspension in as little time as possible.

\end{document}